\documentclass[12pt,a4paper,nofootinbib,superscriptaddress]{revtex4}

\usepackage{ulem}
\usepackage{color}
\usepackage{graphicx}
\usepackage{hyperref}   
\usepackage{appendix}
\usepackage{epsfig}
\usepackage{epstopdf}
\usepackage{amsmath}
\usepackage{subfig}
\usepackage{comment}
\usepackage[dvipsnames]{xcolor}
\usepackage{epsfig}
\usepackage{appendix}
\usepackage{graphics}
\usepackage{graphicx}
\usepackage{color}
\newcommand{\jpsi}{J / \psi}

\newcommand{\old}[1]{}

\newcommand{\be}{\begin{equation}}
\newcommand{\ee}{\end{equation}}
\newcommand{\ba}{\begin{eqnarray}}
\newcommand{\ea}{\end{eqnarray}}
\newcommand{\bi}{\begin{itemize}}
\newcommand{\ei}{\end{itemize}}

\newcommand{\nn}{\nonumber\\}

\begin{document}
\title{Quarkonia dissociation at finite magnetic field in the presence of momentum anisotropy}
\author{Indrani Nilima}
\email{nilima.ism@gmail.com}
\affiliation{Department of Physics, Institute of Science, Banaras Hindu University (BHU), Varanasi, 221005, India}
\author{Mujeeb Hasan}
	\email{mhasan@lords.ac.in}
\affiliation {Lords Institute of Engineering and Technology, Hyderabad- 500091, Telangana, India}	
\author{B. K. Singh}
\email{bksingh@bhu.ac.in}
\affiliation{Discipline of Natural Sciences, PDPM Indian Institute of InformationTechnology Design and Manufacturing, Jabalpur, 482005, India}
\affiliation{Department of Physics, Institute of Science, Banaras Hindu University (BHU), Varanasi, 221005, India}
\author{Mohammad Yousuf Jamal}
	\email{mohammad@iitgoa.ac.in}
\affiliation {School of Physical Sciences, Indian Institute of Technology Goa, Ponda-403401, Goa, India}
\begin{abstract}
   
In this study, we investigate the potential of heavy quarkonia within a magnetized hot QGP medium having finite momentum anisotropy. The phenomenon of inverse magnetic catalysis is introduced into the system, influencing the magnetic field-modified Debye mass and thereby altering the effective quark masses. Concurrently, the impact of momentum anisotropy in the medium is considered that influence the particle distribution in the medium. The thermal decay width and dissociation temperature of quarkonium states, specifically the $1S$ and $2S$ states of charmonium and bottomonium, are computed. Our results reveal that both momentum anisotropy and the inverse magnetic catalysis effects play a significant role in modifying the thermal decay width and dissociation temperature of these heavy quarkonia states.
\end{abstract}
\maketitle

\section{Introduction}
The investigation of heavy quarkonia represents a fundamental avenue for probing the deconfining nature of strongly interacting matter emerging from heavy ion collisions. The hypothesized phase, known as quark-gluon plasma (QGP), resembling conditions post the early universe, is being explored through ongoing ultra-relativistic heavy ion collision (URHIC) experiments at the Relativistic Heavy Ion Collider (RHIC) and Large Hadron Collider (LHC) \cite{Kharzeev:2007jp,Skokov:2009qp,Shovkovy,Muller:PRD89'2013}. These collisions generate remarkably intense magnetic fields, reaching magnitudes estimated at $eB \sim m_{\pi}^2 \equiv 10^{18}$ Gauss at RHIC and $eB \sim 15m_{\pi}^2 \equiv 1.5 \times 10^{19}$ Gauss, owing to the high relative velocities of spectator quarks in non-central events \cite{Kharzeev:2007jp,Skokov:2009qp}.
Initially, there was limited attention given to the impact of magnetic fields on thermal QCD mediums due to the perceived short lifespan of these fields in influencing heavy-ion collision phenomenology. However, recent theoretical predictions suggest that a thermal medium might emerge simultaneously with the production of the magnetic field, resulting in an extended lifespan due to the medium's finite electrical conductivity \cite{Tuchin:2010gx,McLerran:2013hla}. While the field weakens over time, its decay within the medium is slower than in the vacuum. This scenario motivates a comprehensive re-examination of QGP phenomenology in the presence of strong magnetic fields in conjunction with heavy quark phenomenology within an anisotropic medium.
Notably, the inherent translation invariance of space is disrupted by the magnetic field, leading to an expected absence of anisotropy in the potential concerning the field's direction. In our approach, the emergence of anisotropy in the potential within the coordinate space originates from the manifested momentum anisotropy associated with the direction of anisotropy in the distribution function.

Recent studies have examined the impact of magnetic fields on various QCD phenomena \cite{Alford:2013jva,Fukushima:2015wck,Das:2016cwd,Singh:2017nfa,Hasan:2017fmf,Kurain:2019,Bandyopadhyay:2021zlm,Nilima:2022tmz}, uncovering phenomena such as magnetic catalysis, inverse magnetic catalysis, axial magnetic effects, and chiral magnetic effects. These investigations prompt a deeper exploration of the interplay between strong magnetic fields and heavy quark phenomenology within an anisotropic medium. In continuation of our recent work on the dissociation of the heavy quark potential in the presence of finite magnetic fields \cite{Nilima:2022tmz}, this study focuses on introducing local momentum anisotropy to investigate the dissociation temperature of heavy quarkonia. It incorporates the effects of inverse magnetic catalysis through the magnetic field-modified Debye mass. Recent lattice quantum chromodynamics (LQCD) calculations have suggested coexisting magnetic catalysis and inverse magnetic catalysis at non-zero QCD vacuum temperatures and magnetic fields \cite{Bali1,Bali2}. This work integrates the crucial inverse magnetic catalysis effect derived from constituent quark masses obtained through lattice QCD simulations.
The primary goal here is to explore the inverse magnetic catalysis (IMC) effect on the constituent quark masses within the heavy quark potential in an anisotropic medium. To achieve this, we establish the standard formalism of the heavy quark potential in a magnetic field as a foundation for our investigation \cite{Agotiya:2008ie,Thakur:2012eb,Thakur:2013nia,Kakade:2015xua,Agotiya:2016bqr,Jamal:2023ncn}. Subsequently, we derive the real and imaginary components of an anisotropic heavy quark potential in a magnetic field, yielding modified binding energies and thermal widths of quarkonium states. Finally, employing the magnetic field-modified heavy quark potential within an anisotropic medium, we scrutinize the dissociation of charmonium and bottomonium states.

The paper is organized as follows. In section~\ref{sec2},  we will discuss the basic formalism of our present work, which includes discussions about the real and imaginary parts of the heavy quark potential, decay width, binding energy, and the Debye screening mass considering  IMC and momentum anisotropy effects. In section~\ref{sec3}, we will show our results for the same, as well as find out the dissociation temperatures and discuss their magnetic field dependence. Finally, in section~\ref{sec4}, we shall conclude the present work.

\section{Formalism}
\label{sec2}
The quark-antiquark interaction in the quarkonia state is characterized by the Cornell potential~\cite{Eichten:1978tg, Eichten:1979ms, Sebastian:2022sga}. This potential comprises both Coulombic and string components, denoted as:

\begin{equation}
V(r) = -\frac{\alpha}{r}+\sigma r~.
\label{eq:cor}
\end{equation}

Here, $r$ represents the effective radius of the heavy quark and antiquark, $\alpha$ is the strong coupling constant given by $({\alpha}=\frac{g_s^2C_F}{4\pi}; C_F=4/3)$, and $\sigma$ is the string tension. 
We introduce anisotropy at the particle phase space distribution level in our formalism using the method outlined in Refs.~\cite{Romatschke:2003ms, Carrington:2014bla, Jamal:2017dqs, Jamal:2020hpy, Kumar:2017bja}, we obtain anisotropic distribution functions from isotropic ones by rescaling (stretching and squeezing) in one direction in momentum space:

\begin{equation}
f({\mathbf{k}})\rightarrow f_{\xi}({\mathbf{k}}) = C_{\xi}~f(\sqrt{{\bf k}^{2} + \xi({\bf k}\cdot{\bf \hat{n}})^{2}}),
\label{aniso_distr}
\end{equation}

Here, $f({\mathbf{k}})$ is the isotropic distribution function of the particle, and ${\mathbf{\hat{n}}}$ is a unit vector indicating the direction of momentum anisotropy. The parameter $\xi$ quantifies the anisotropic strength in the medium, describing the degree of stretching ($-1<\xi<0$, or prolate form) and squeezing ($\xi > 0$, or oblate form) in the ${\bf \hat{n}}$ direction. $C_{\xi}$ is the normalization constant that can be obtained either by keeping particle number or the debye mass intact in both the medium.

The magnetic field-dependent Cornell potential in the presence of an anisotropic hot QCD medium is obtained by dividing it with the dielectric permittivity $\epsilon(k,T,eB,{\xi})$ in Fourier space. This permittivity encapsulates the medium information such as temperature, anisotropy, and magnetic field. The modified Cornell potential is thus obtained as,

\begin{equation}
{\widetilde{V}}(k,T,eB,{\xi})=\frac{{{ V}}(k)}{\epsilon(k,T,eB,{\xi})}~,
\end{equation}

and by taking the inverse Fourier transform, we have

\begin{equation}
V(r,T,eB,{\xi})= \int \frac{d^3\mathbf{k}}{(2\pi)^{3/2}}(e^{i\mathbf{k} \cdot \mathbf{r}}-1)\widetilde{V}(k,T,eB,{\xi})~.
\label{eq:V}
\end{equation}

The dielectric permittivity of the medium is associated with the longitudinal part of the gluon self-energy~\cite{schneider} and can be obtained in Fourier space from the $\omega\to 0$ limit of the temporal component of the medium-dependent effective gluon propagator ($\Delta^{00}$) as:

\begin{equation}
\epsilon^{-1}({\bf k},T,eB,{\xi}) = -\lim_{\omega \to 0}k^2\Delta^{00}(\omega,{\bf k},T,eB,{\xi}).
\label{eq:eps}
\end{equation}

To obtain the real part of dielectric permittivity, we derive the real part of the propagator given as,

\ba
  Re[\Delta^{00}]({\omega = 0,{\bf k},T,eB,{\xi}}) &=&\frac{-1}{k^2+m_{D}^2(T,eB)}+\xi\Big(\frac{1}{3\{k^2+m_{D}^2 (T,eB)\}}\nn
  &-&\frac{m_{D}^2(T,eB)(3\cos{2\theta_n}-1)}{6\{k^2+m_{D}^2(T,eB)\}^2}\Big).
\label{eq:Re_delta}
\ea

In a similar fashion, for the imaginary part of dielectric permittivity, the imaginary part of the propagator is determined as follows,
\begin{align}
\text{Im}[\Delta^{00}]({\omega = 0,{\bf k},T,eB,{\xi}}) =& \pi{m_{D}^2(T,eB)}T \bigg(\frac{-1}{k\{k^2+m_{D}^2(T,eB)\}} 
+ \xi\Big(\frac{-1}{3k\{k^2+m_{D}^2(T,eB)\}^2} \nn &+ \frac{3\sin^{2}{\theta_n}}{4k\{k^2+m_{D}^2(T,eB)\}^2} 
- \frac{2m_{D}^2(T,eB)\big(3\sin^{2}({\theta_n})-1\big)}{3k\{k^2+m_{D}^2(T,eB)\}^3}\Big)\bigg),
\label{eq:Im_delta}
\end{align}

Here, $\theta_n$ represents the angle between ${\bf k}$ and $\bf{ \hat n}$. Additionally, $\hat{s}=rm_D$ and $m_D\equiv m_D(T,eB)$ is the Debye mass~\cite{Nilima:2022tmz}, characterizing the IMC effects through the medium-dependent constituent quark mass $M_f(T, eB)$. This mass is derived from lattice QCD-predicted values for normalized quark condensates $\langle q \bar{q}\rangle_f (T,eB)$ at finite $eB$~\cite{Bali1,Bali2}:
\begin{align}
M_f(T,eB) = M_f(T=0,eB=0)\times \langle q{\bar q}\rangle_f(T,eB) + m_f, 
\end{align}
where $m_f$ is the light quark bare mass. In the limit $T \gg m_f$, we have 
\begin{align}
M_f(T,eB) \approx M_f(T=0,eB=0)\times \langle q{\bar q}\rangle_f(T,eB).
\label{M_TB}
\end{align}

The dispersion relation, now modified to incorporate the effective constituent mass in a magnetized medium, is expressed as:
\begin{equation}
\bar{E}_f^l = \sqrt{k_z^2+2l|q_f eB|+M_f(T,eB)^2}.
\label{disp_imc}
\end{equation}

Finally, employing the lattice QCD-inspired modified dispersion relation from Eq.~(\ref{disp_imc}), the Debye screening mass is evaluated as:
\begin{eqnarray}\label{6}
m_{D}^{2} &=& 4\pi\alpha_s(T) T^2 \frac{N_c}{3} + \sum_f \dfrac{4\pi\alpha_s(T)|q_feB|}{\pi^2T} \nn
&\times& \int_{0}^{\infty} dk_z \sum_{l=0}^{\infty}(2-\delta_{l0})~{f}^l_q(\bar{E}_f^l)~\left(1-{f}^l_q(\bar{E}_f^l)\right),
\end{eqnarray}

where $\alpha_s(T)$ is the temperature-dependent one-loop running coupling~\cite{Haque:2018eph}:
\begin{eqnarray}
\label{as}
g_s^2(T) = 4\pi\alpha_{s}(T) = \frac{24 \pi^2}{\left(11N_c-2 N_{f}\right)\ln \left(\frac{2\pi T}{\Lambda_{\overline{\rm MS}}}\right)},
\end{eqnarray}
with $\Lambda_{\overline{\rm MS}}$ representing the $\overline{\rm MS}$ renormalization scale.
Next, considering the short-distance limit $ \hat{s} \ll 1$ ~\cite{Agotiya:2016bqr,V_mD1,V_mD2}, we obtained of the real and imaginary parts of the potential given as,

\ba
Re[V(\hat{s},T,eB,\xi)] &=&\frac{ \hat{s}~ \sigma }{m_D}\left(1+\frac{\xi }{3}\right)-\frac{\alpha~  m_D}{\hat{s}} \bigg(1+\frac{\hat{s}^2}{2}
+\xi  \left(\frac{1}{3}+\frac{\hat{s}^2}{16}\left(\frac{1}{3}+ \cos \left(2 \theta _r\right)\right)\right)\bigg),
\label{eq:realv}
\ea

\ba
\text{Im}[V (\hat{s},T,eB,\theta_r,\xi)]&=&\frac{\alpha ~ \hat{s}^2~ T}{3} \Big\{\frac{ \xi }{60} (7-9 \cos 2 \theta_r)-1\Big\}\log \left(\frac{1}{\hat{s}}\right)\nn
&+&\frac{\hat{s}^4 ~\sigma ~ T}{m_D^2}\Big\{\frac{\xi}{35}  \left(\frac{1}{9}-\frac{1}{4} \cos 2 \theta_r \right)
-\frac{1}{30}\Big\}\log \left(\frac{1}{\hat{s}}\right),
 \label{eq:imagv}
 \ea

where $\theta_r$ is the angle between $\bf{r}$ and $\bf{ \hat n}$. With this simplification, the information of $T$ and $eB$ in the heavy quark potential enters solely through the Debye mass. Next, we discuss the binding energy and thermal width of heavy quarkonia. For the binding energy calculation under small anisotropy, we solve the Schr\"odinger equation using the real part of the potential as done in ~\cite{Margotta:2011ta, Strickland:2011aa, Thakur:2013nia}. It is important to note that we favor the real part over the imaginary part due to its substantial magnitude (clearly depicted in Figs.\ref{repotN} to \ref{impot1} for different sets of parameters). The real part of the binding energy ${Re}[BE]$ is expressed as:

\begin{equation}
{\text{Re}[BE(T,eB)]} = \bigg( \frac{m_Q\sigma^2 }{m_{D}^4 n^{2}} + \alpha m_{D} +\frac{\xi}{3}\Big(\alpha m_{D}+ \frac{3m_Q\sigma^2 }{m_{D}^4 n^{2}}\Big)\bigg),
\label{realp}
\end{equation}

The imaginary part of the potential give rise to the thermal width ($\Gamma$). For the various quarkonia state, employing the imaginary part of the potential, $\text{Im}~ V (\hat{s},T,B,\theta_r,\xi)$, the thermal width is computed as~\cite{Singh:2023zxu, Agotiya:2016bqr}:

\begin{align}
\Gamma(T,eB) = - \int d^3{\bf{r}}\, \left|\Psi({{r}})\right|^2{\rm{Im}}~\ V({\hat{s};T,B, \theta_r, \xi})~,
\label{Gamma}
\end{align}

It is essential to highlight that we choose the Coulombic wave function $\Psi(r)$ due to its long-range Coulombic tail, dominant at higher temperatures. We consider both the Coulombic wave functions for the ground state ($1s$, $J/\psi$ and $\Upsilon$) and the first excited state ($2s$, $\psi'$ and $\Upsilon'$), respectively given as:

\begin{align}
\Psi_{1s}(r) &= \frac{1}{\sqrt{\pi B_r^3}}e^{-\frac{r}{B_r}},\\
\Psi_{2s}(r) &= \frac{1}{2\sqrt{2\pi B_r^3}}\left(2-\frac{r}{B_r}\right)e^{-\frac{r}{2B_r}},
\label{psi}
\end{align}

Where $B_r$=$2/(m_Q\alpha_s)$ represents the Bohr radius of the quarkonia system. Utilizing Eq.\ref{Gamma}, the thermal width for $1s$ and $2s$ states appears as follows, respectively:

\begin{align}
 \Gamma_{1s}(T,eB) &=  T\bigg(\frac{4}{\alpha_s ~m_Q^2}+\frac{12\sigma}{{\alpha_s}^4~m_Q^4}\bigg) 
 \bigg(1-\frac{\xi}{6}\bigg)m_D^2 \log\left(\frac{m_D}{\alpha_s~ m_Q}\right),
 \label{Gamma1sl}
 \end{align} 
\begin{align}
 \Gamma_{2s}(T,eB) &=  \frac{8 ~m_D^2 T}{{\alpha_s}^4~ m_Q^4}\Big(1-\frac{\xi}{6}\Big)\bigg(7 {\alpha_s} ^3 m_Q^2+192 ~\sigma \bigg)\log \left(\frac{2~ m_D}{{\alpha_s} ~m_Q}\right).
 \label{Gamma2s}
 \end{align}
 
Now, having the binding energy and the thermal dissociation width for both ground and first excited states of charmonium and bottomonium states (i.e., $\jpsi/\psi'$ and $\Upsilon/\Upsilon'$), respectively, we can determine the dissociation temperature for the quarkonium states using the criteria $\Gamma_{n}(T,eB) = 2\times B. E_{n}(T,eB) $~\cite{Mocsy:2007jz} for $n=1$ and $2$.


\section{Results}
\label{sec3}
\begin{figure} 
	\begin{center}
		\includegraphics[width=7.5cm]{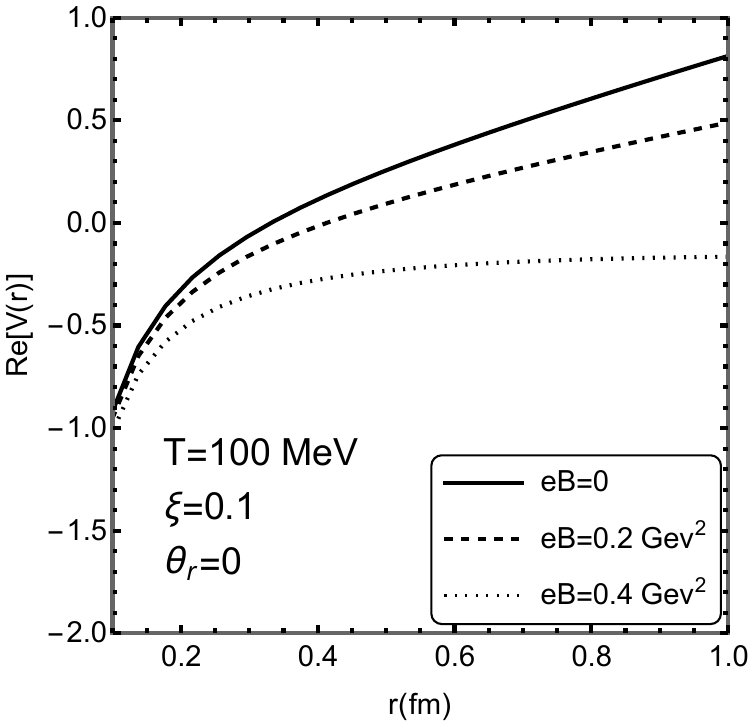}
  \hspace{5mm}
		\includegraphics[width=7.5cm]{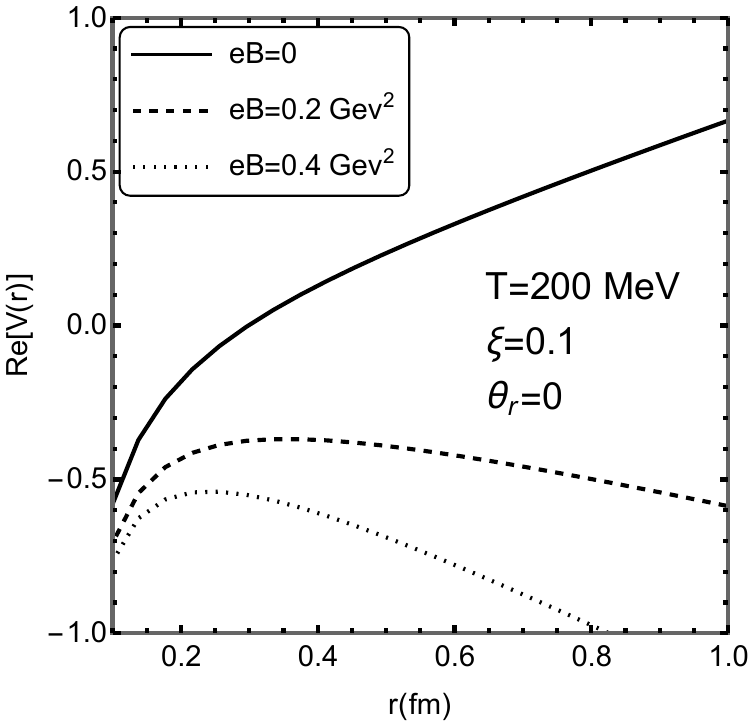}
		\caption{Variation of the real part of potential with separation distance $r$ between $Q\bar{Q}$ for three different values of the magnetic field at $\theta_r$ =0 and with fixed temperature T=100 MeV (left) and T=200 MeV (right)  at $\xi = 0.1$.}
		\label{repotN}
	\end{center}
\end{figure}

\begin{figure} 
	\begin{center}
		\includegraphics[width=7.5cm]{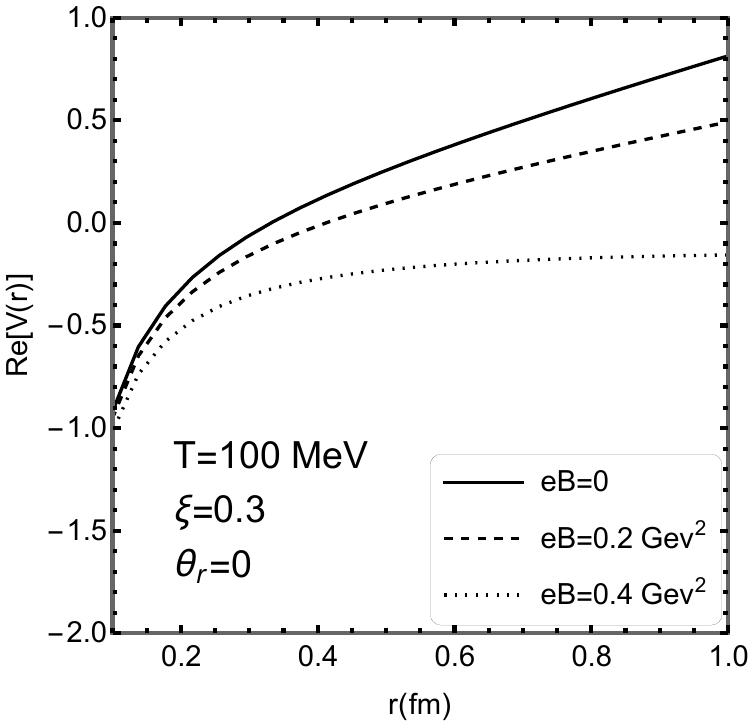}
  \hspace{5mm}
		\includegraphics[width=7.5cm]{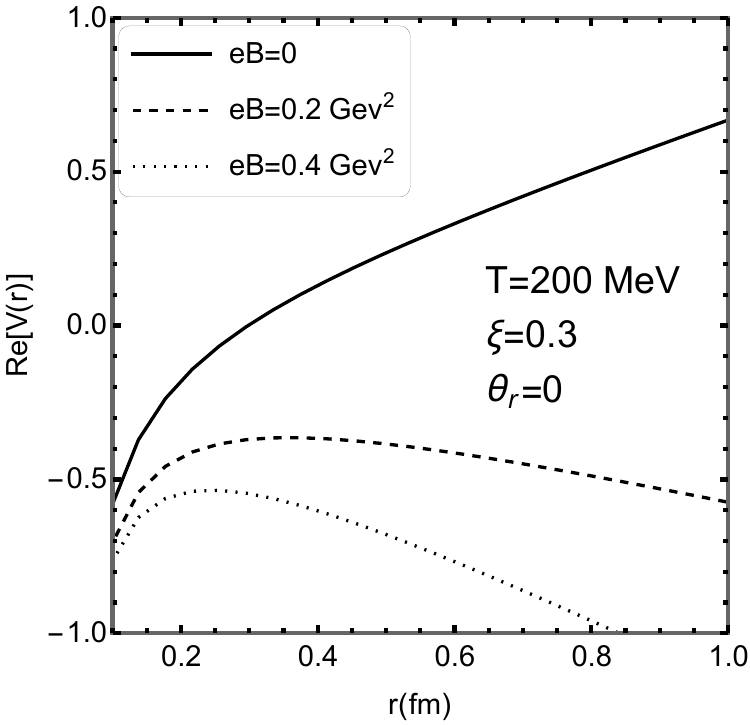}
		\caption{Variation of the real part of potential with separation distance $r$ between $Q\bar{Q}$ for three different values of the magnetic field at $\theta_r$ =0 and with fixed temperature T=100 MeV (left) and T=200 MeV (right)  at $\xi = 0.3$.}
		\label{repot}
	\end{center}
\end{figure}

\begin{figure} 
\begin{center}
		\includegraphics[width=7.5cm]{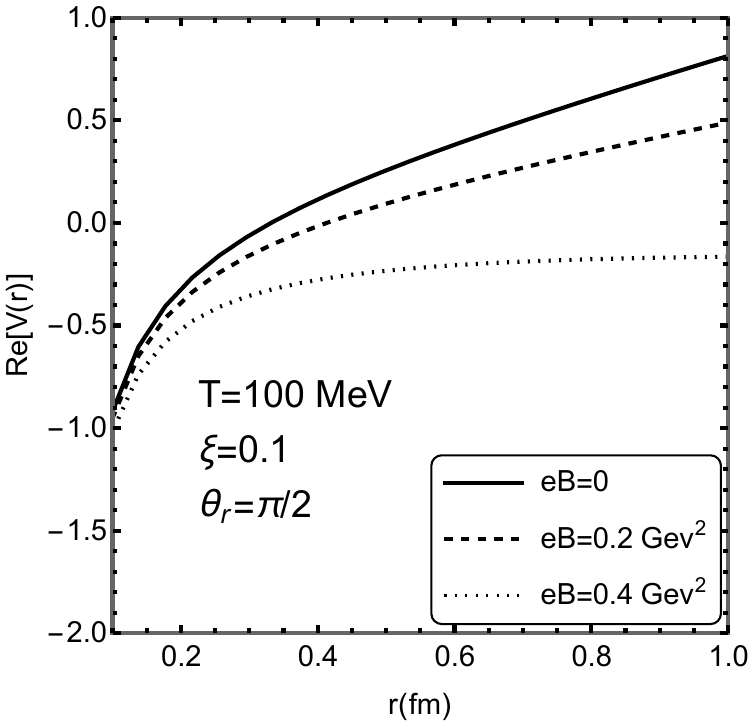}
  \hspace{5mm}
		\includegraphics[width=7.5cm]{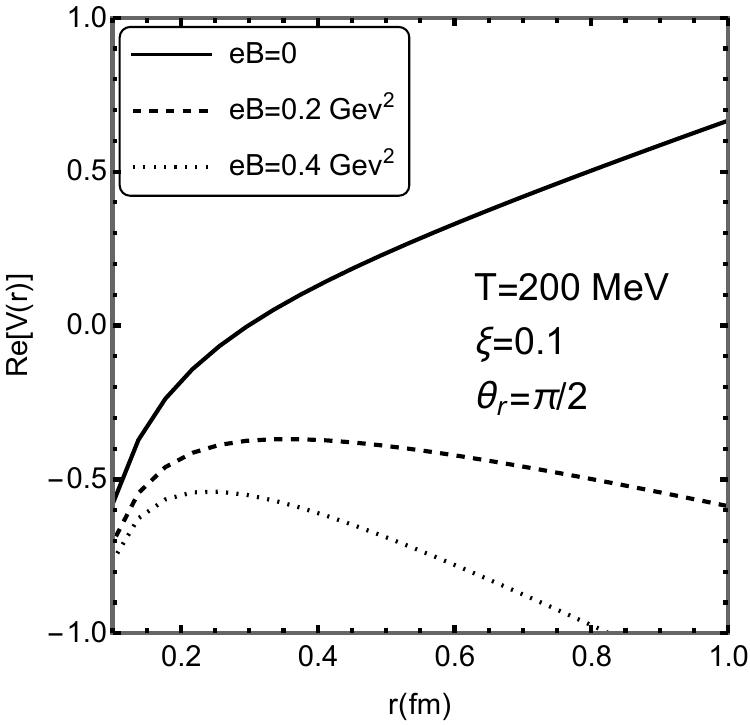}
		\caption{Variation of the real part of potential with separation distance $r$ between $Q\bar{Q}$ for three different values of the magnetic field at $\theta_r =\pi/2$ and with fixed temperature T=100 MeV (left) and T=200 MeV (right) at $\xi = 0.1$.}
		\label{repot1N}
	\end{center}
\end{figure}
\begin{figure} 
	\begin{center}
		\includegraphics[width=7.5cm]{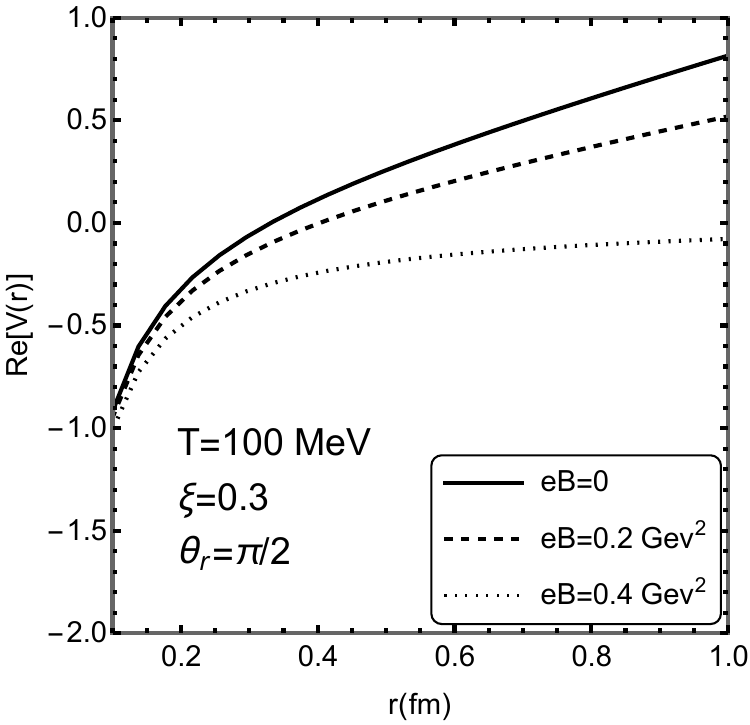}
  \hspace{5mm}
		\includegraphics[width=7.5cm]{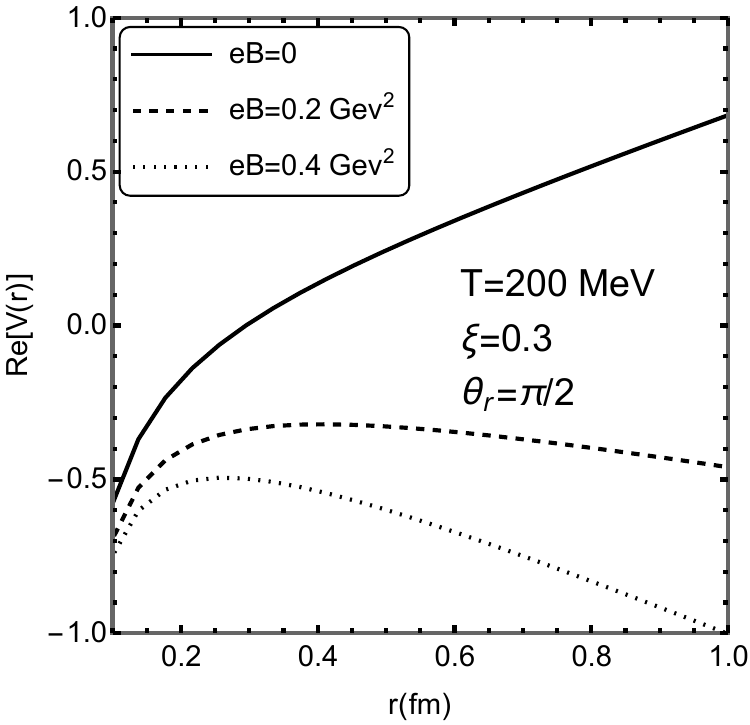}
		\caption{Variation of the real part of potential with separation distance $r$ between $Q\bar{Q}$ for three different values of the magnetic field at $\theta_r =\pi/2$ and with fixed temperature T=100 MeV (left) and T=200 MeV (right) at $\xi = 0.3$.}
		\label{repot1}
	\end{center}
\end{figure}
\begin{figure}  
	\begin{center}
		\includegraphics[width=7.5cm]{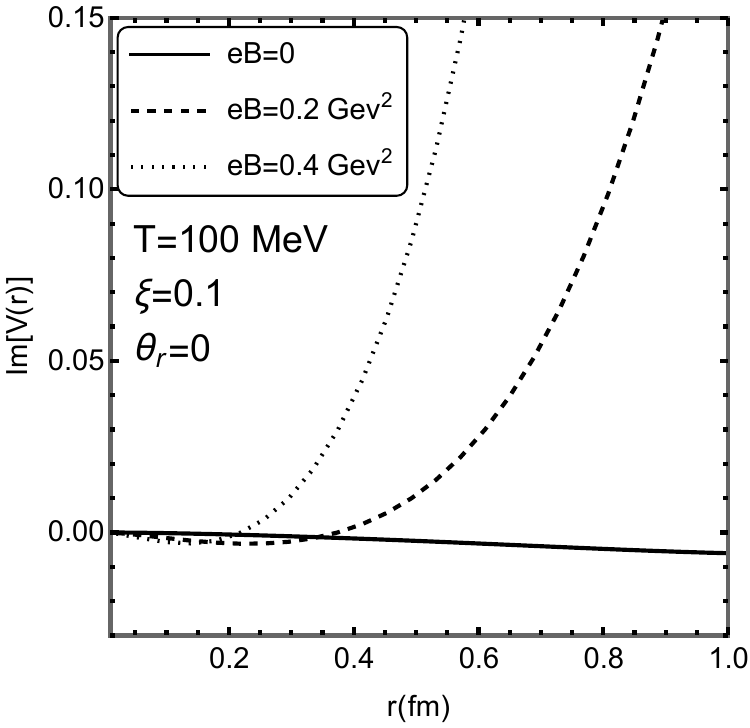}
  \hspace{5mm}
		\includegraphics[width=7.5cm]{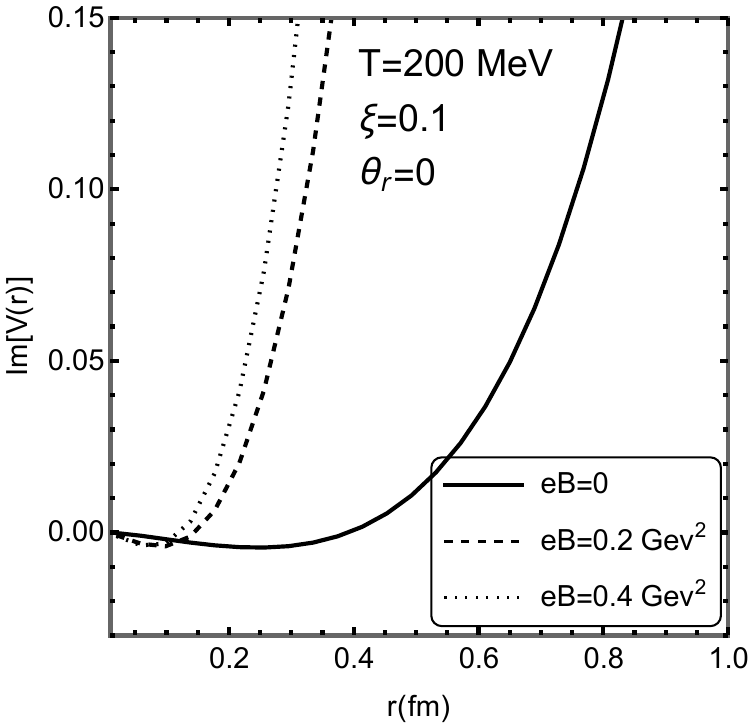}
		\caption{Variation of the imaginary part of potential with separation distance $r$ between $Q\bar{Q}$ for various values of magnetic field at $\theta_r$ =0 with temperature T=100 MeV (left) and T=200 MeV (right)  at $\xi = 0.1$.}
		\label{impotN}
	\end{center}
\end{figure}
\begin{figure}  
	\begin{center}
		\includegraphics[width=7.5cm]{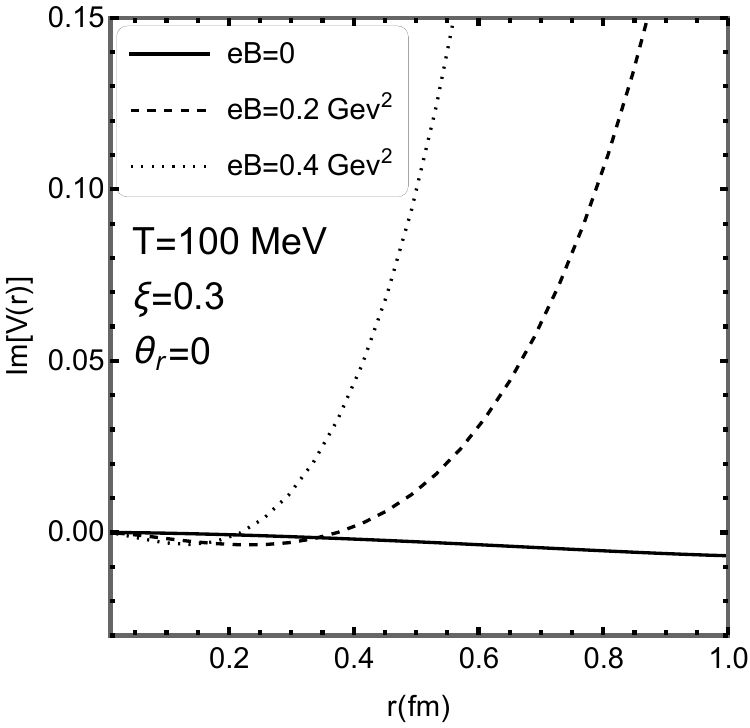}
  \hspace{5mm}
		\includegraphics[width=7.5cm]{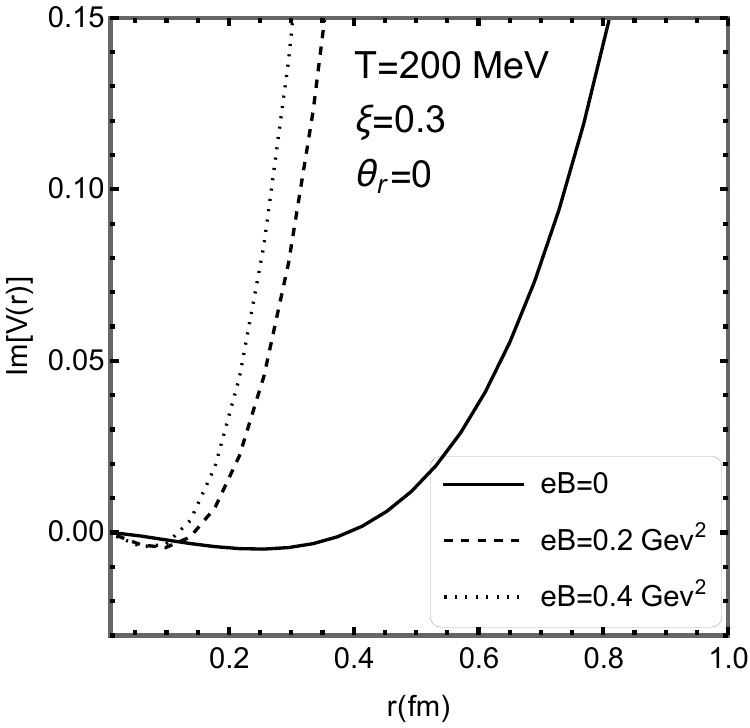}
		\caption{Variation of the imaginary part of potential with separation distance $r$ between $Q\bar{Q}$ for various values of magnetic field at $\theta_r$ =0 with temperature T=100 MeV (left) and T=200 MeV (right)  at $\xi = 0.3$.}
		\label{impot}
	\end{center}
\end{figure}

\begin{figure}  
	\begin{center}
		\includegraphics[width=7.5cm]{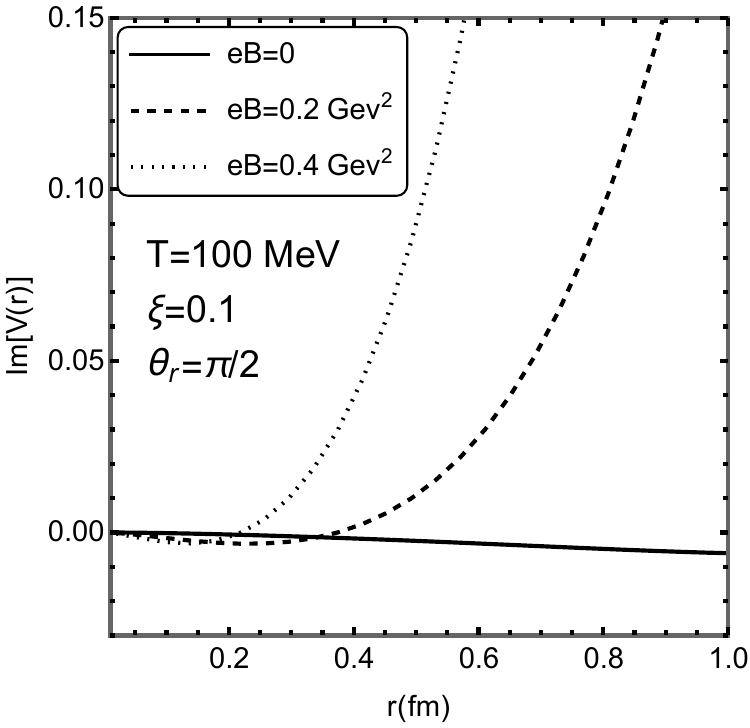}
  \hspace{5mm}
		\includegraphics[width=7.5cm]{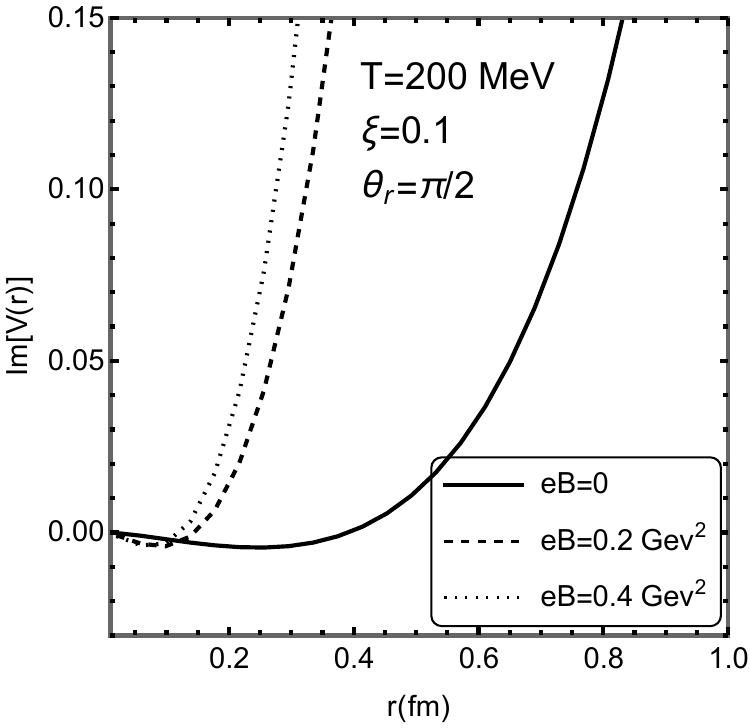}
		\caption{Variation of the imaginary part of potential with separation distance $r$ between $Q\bar{Q}$ for various values of magnetic field at $\theta_r =\pi/2$ with temperature T=100 MeV (left) and T=200 MeV (right)  at $\xi = 0.1$.}
		\label{impot1N}
	\end{center}
\end{figure}
\begin{figure}  
	\begin{center}
		\includegraphics[width=7.5cm]{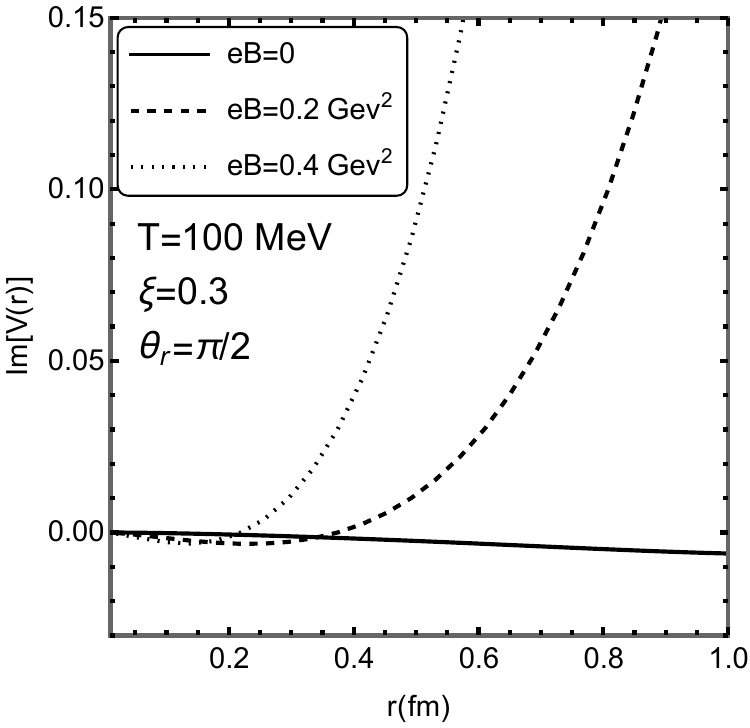}
  \hspace{5mm}
		\includegraphics[width=7.5cm]{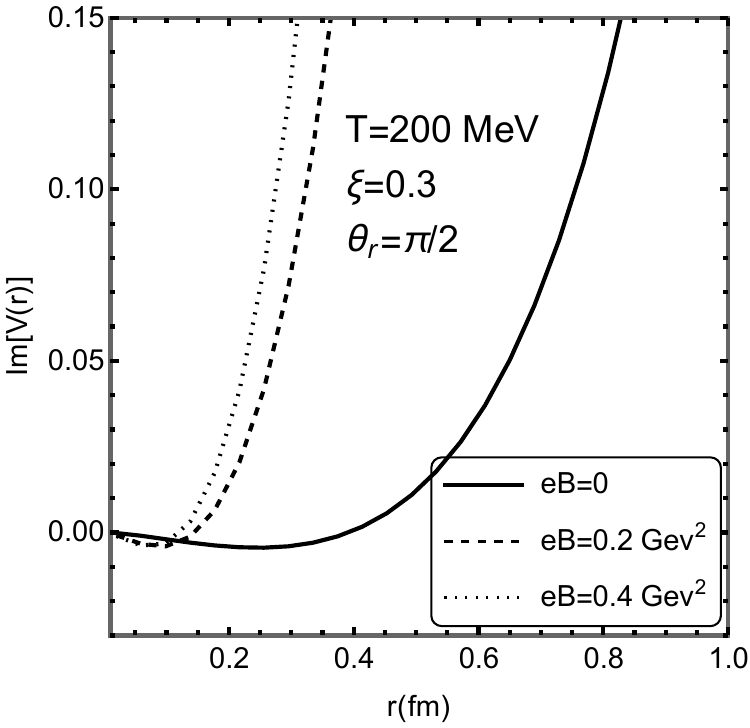}
		\caption{Variation of the imaginary part of potential with separation distance $r$ between $Q\bar{Q}$ for various values of magnetic field at $\theta_r =\pi/2$ with temperature T=100 MeV (left) and T=200 MeV (right)  at $\xi = 0.3$.}
		\label{impot1}
	\end{center}
\end{figure}
\begin{figure} 
	\begin{center}
		\includegraphics[width=7.5cm]{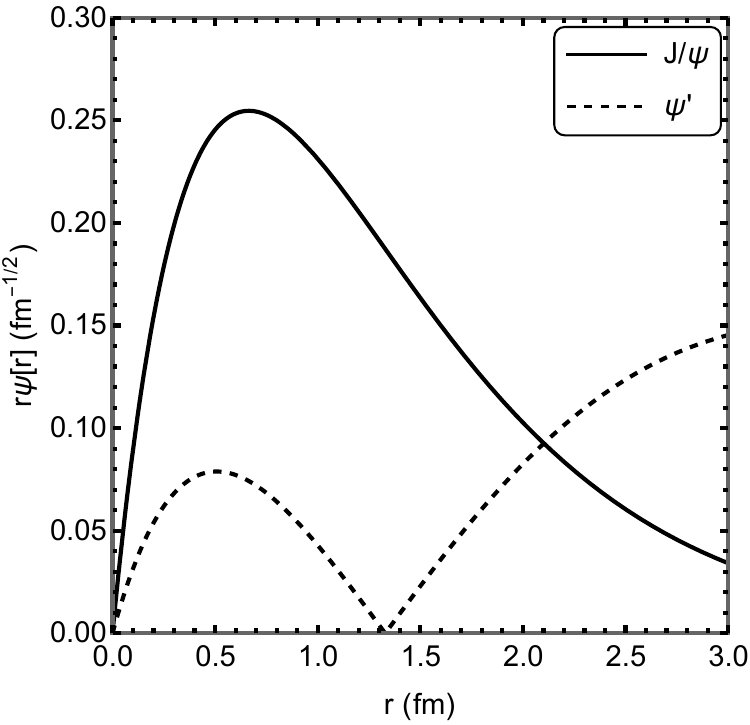} 
  \hspace{5mm}
            \includegraphics[width=7.5cm]{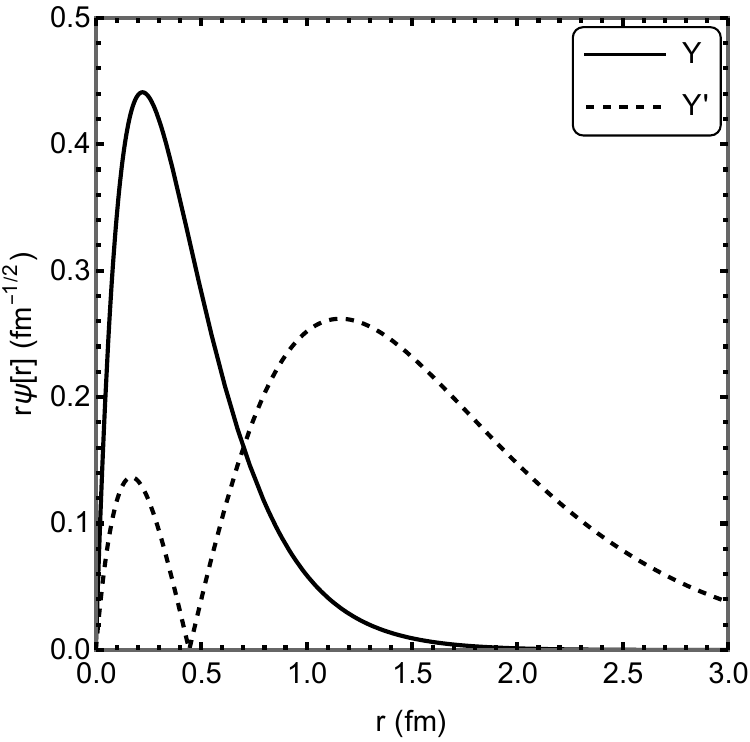} 
		\caption{The variation of the wavefunction of different charmonia (left) and bottomonium (right)  states with respect to $r$ at T=200 MeV. The solid curve represents the 1S state, and the dotted curve is for the 2S state. }
		\label{wfjpsi}
	\end{center}
\end{figure}

%
\begin{figure} 
	\begin{center}
		\includegraphics[width=7.5cm]{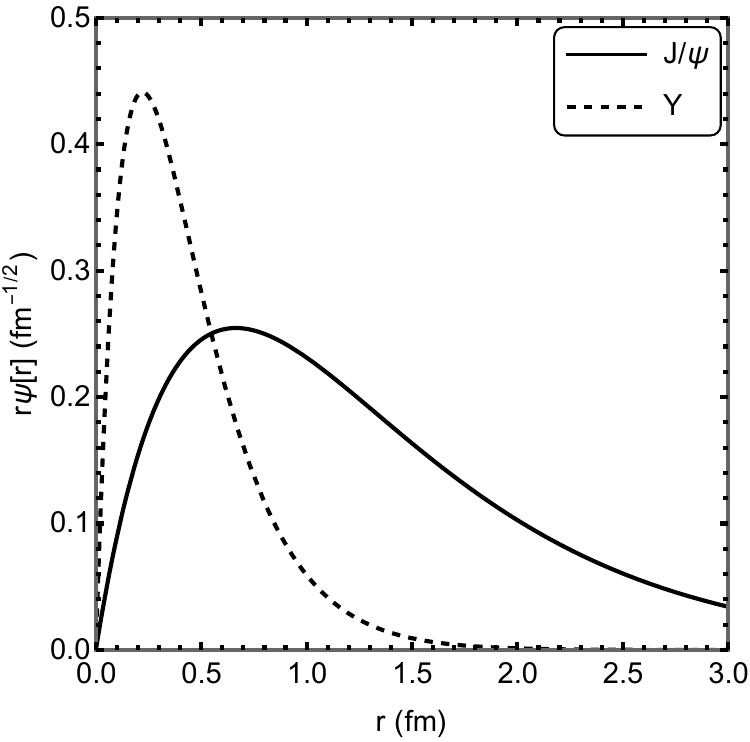} 
  \hspace{5mm} 
  \includegraphics[width=7.5cm]{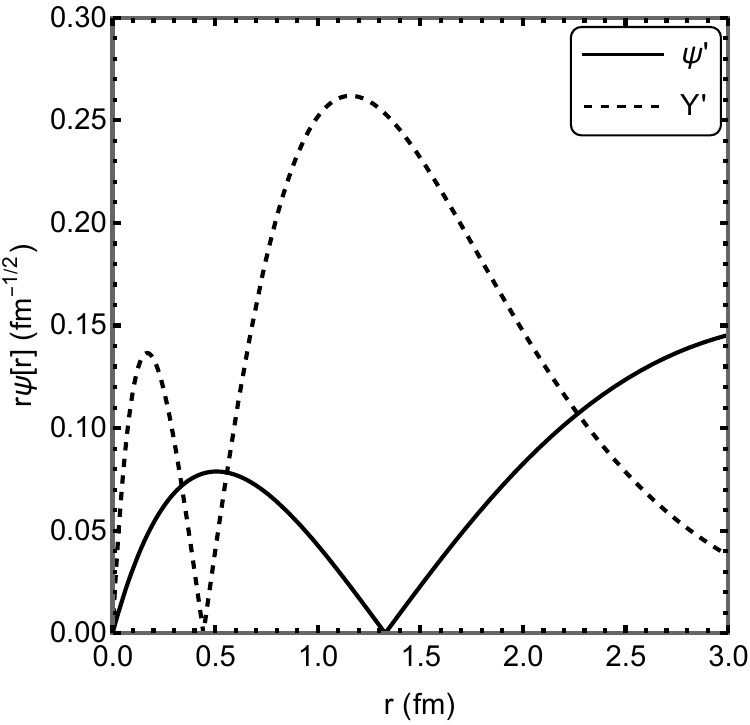} 
		 \caption{The variation of the wavefunction with respect to $r$ for 1s state (left) and 2s state (right) at T=200 MeV. The solid curve represents the charmonium state, and the dotted curve shows the bottomonium state. }
		\label{wfjpsip} 
	\end{center}
\end{figure}
In this section, we present significant findings pertaining to the magnetic field-dependent modified heavy quarkonia potential, incorporating the IMC-based quark condensate and accounting for momentum anisotropy. The analysis encompasses an exploration of the impact of weak anisotropy within the hot QCD plasma under specific conditions, $\xi=0.1$ and $\xi=0.3$. The other parameters are fixed as $N_f=2$ quark flavors and $N_c=3$ colors, with $\Lambda_{\overline{\rm MS}} = 176$ MeV ~\cite{Haque:2014rua}, and a string tension value of $\sigma = 0.184$ ~GeV$^2$ as indicated in \cite{Jamal:2018mog}.

 Figures \ref{repotN} and \ref{repot} depict the real part of the potential as a function of the separation distance ($ r $) between the $Q\bar{Q}$ pair under different magnetic field strengths ($eB= 0, 0.2$ GeV$^2$, and $0.4$ GeV$^2$). These results are showcased for $\xi=0.1$ and $0.3$, specifically exploring the parallel case ($\theta_r =0$) across two temperatures: $T= 100 $ MeV (left panel) and $T= 200$ MeV (right panel). Corresponding plots for the perpendicular case ($\theta_r = \pi/2$) are displayed in Figs. \ref{repot1N} and \ref{repot1}. For the case of $\xi=0.3$, there is a discernible increment in the values compared to $\xi=0.1$. In both parallel and perpendicular configurations ($\theta_r = 0$ and $\theta_r = \pi/2$), the real potential initiates from a negative value and sharply rises towards zero, exhibiting a more pronounced behavior at $T=100$ MeV than at $T=200$ MeV for both anisotropy values. This observation suggests a tendency for the potential to stabilize at lower values with increasing temperature. Moreover, it is noteworthy that the screening effect is more pronounced at higher temperatures ($T= 200 $ MeV) compared to lower temperatures ($T= 100 $ MeV) for both $\xi=0.1$ and $0.3$, irrespective of the orientation (parallel or perpendicular). Additionally, there's a subtle sensitivity to anisotropy, where a slightly larger separation is observed for $\theta_r = \pi/2$ compared to $\theta_r = 0$. Concurrently, the presence of the magnetic field serves to diminish the magnitude of the real part of the potential, evident across both anisotropy values.

Employing the identical parameters, the imaginary part of the potential is illustrated in Figs. \ref{impotN} and \ref{impot} for $\theta_r =0$, corresponding to $\xi=0.1$ and $\xi=0.3$ respectively. Similarly, Figs. \ref{impot1N} and \ref{impot1} present the outcomes for $\theta_r = \pi/2$ with $\xi=0.1$ and $\xi=0.3$. Across both orientations, the imaginary component consistently amplifies as the magnetic field strength escalates, whether parallel or perpendicular. In terms of magnitude, this imaginary part exhibits greater prominence at higher temperatures ($T= 200 $ MeV) compared to lower temperatures ($T= 100 $ MeV). This difference arises due to the dependency on temperature and the Debye mass associated with the magnetic field. Notably, the distinction between the magnetic field-dependent imaginary potentials is more conspicuous at elevated temperatures ($T= 200 $ MeV) than at lower temperatures ($T= 100 $ MeV), observed consistently for both $\theta_r = 0$ and $\theta_r = \pi/2$. Moreover, the presence of the magnetic field suppresses the magnitude of the imaginary potential.

Comparing with our earlier work~\cite{Nilima:2022tmz}, where we previously explored the variation of both real and imaginary parts of the potential concerning the separation distance ($r$) between $Q\bar{Q}$ pairs for distinct magnetic field values at temperatures $T=100 $ MeV and $T=200 $ MeV, excluding momentum anisotropy in the medium, reveals intriguing insights. The exponential decay with distance becomes notably more pronounced at higher temperatures ($T= 200 $ MeV) compared to lower temperatures ($T= 100 $ MeV). Similarly, transitioning from zero to non-zero magnetic field values and increasing their magnitudes accentuates this dominance in the exponential behavior.

Fig. (~\ref{wfjpsi}) shows the variation of the radial part of the eigen wavefunction for $1s$ and $2s$ states of charmonium (left panel) and bottomonium (right panel) with respect to $r$. Noticeable differences are observed in the peak height and peak width of $1s$ state ($J/\psi$ and $\psi'$) compared to $2s$ state ($\Upsilon$, and  $\Upsilon'$) for both charmonium and bottomonium. The stronger binding nature of the $1s$ state ($J/\psi$ and $\Upsilon$) is evident compared to the $2s$ state ($\psi'$, and  $\Upsilon'$). Fig. (~\ref{wfjpsip}) provides a comparison between the eigen wavefunctions of $1s$ state ($J/\psi$ and $\Upsilon$) and $2s$ state ($\psi'$ and  $\Upsilon'$). Consistent with the earlier conclusion, we observe that the binding nature of the $1s$ state ($J/\psi$ and $\Upsilon$) is stronger than that of the $2s$ state ($\psi'$ and  $\Upsilon'$).

In broad terms, the real and imaginary components of the potential undergo modifications influenced by temperature ($T$) and external magnetic field strength ($eB$) in the presence of momentum anisotropy. These results stem from the intricate behavior of Lattice QCD (LQCD) based quark condensates, indicating an enhanced screening effect and a tendency towards dissociation with rising values of $T$ and $eB$. 

This behavior is again due to the aforementioned $T$ and $eB$ dependence of the Debye mass. So, grossly, we notice that real and imaginary parts of heavy quark potential are modified with $T$ and $eB$ due to the non-trivial profile of LQCD-based quark condensates, and these modifications suggest more screening and dissociation with increasing of $T$ and $eB$. They also indicate the possibilities of low dissociation temperatures due to the external magnetic field, which we will explicitly see later.

Extending our analysis, we employ the thermal width expressed in Eqs.~\eqref{Gamma1sl}/\eqref{Gamma2s}, coupled with the binding energy detailed in Eq.~\eqref{realp}, to determine the dissociation temperature ($T_d$) as the point of their intersection. Our findings regarding dissociation temperatures are summarized in Tables \ref{tab:t1} and \ref{tab:t2} for $\xi=0.1$ and $0.3$, respectively. Specifically, at $\xi=0.1$, the $J/\psi$ state undergoes dissociation at $T=192 $ MeV for $eB=0$, $T=123 $ MeV for $eB=0.2$ GeV$^2$, and $T= 95 $ MeV for $eB=0.4$ GeV$^2$. Conversely, the $\Upsilon$ state experiences dissociation at $T= 290 $ MeV for $eB=0$, $T=217 $ MeV for $eB=0.2$ GeV$^2$, and $T= 207 $ MeV for $eB=0.4$ GeV$^2$ in the $\xi=0.1$ case. Notably, the dissociation temperature is the lowest for the $\Upsilon^{'}$ ($2s$-state), whereas it is higher for the $\Upsilon$ ($1s$-state) compared to $J/\psi$ ($1s$-state). This hierarchy persists across all three different magnetic field strengths, i.e., $eB=0$, $eB = 0.2$ GeV$^2$, and $eB=0.4$ GeV$^2$. In all the plots, the intersection points of thermal width and binding energy are found to be smaller for $\xi=0.1$ compared to the $\xi=0.3$ case. Thus, for $\xi=0.3$, we observe slightly higher values of the dissociation temperature for the quarkonium states at $eB=0, 0.2$, and $0.4$ GeV$^2$ compared to the previous scenario.

\begin{table}[tbh]
	\begin{center}
		\begin{tabular}{ |p{4.5cm}||p{2.5cm}|p{3.0cm}|p{3.0cm}|  }
			\hline
State $\downarrow$  Field strength $\rightarrow$  &eB=0 & eB=0.2 GeV$^2$& eB=0.4 GeV$^2$\\
			\hline\hline
			$\jpsi$&192 & 123 & 95 \\
			\hline
			$\Upsilon$&290 & 217 & 207 \\
			\hline
			\hline
			$\Upsilon^{'}$&180 & 111 & 88 \\
			\hline
		\end{tabular}
		\caption{The dissociation temperature($T_D$) for the
			quarkonia states (in units of MeV), when $\Gamma$= 2BE at $\xi$ =0.1} 
		\label{tab:t1}
	\end{center}
\end{table} 

\begin{table}[tbh]
	\begin{center}
		\begin{tabular}{ |p{4.5cm}||p{2.5cm}|p{3.0cm}|p{3.0cm}|  }
			\hline
State $\downarrow$  Field strength $\rightarrow$  &eB=0 & eB=0.2 GeV$^2$& eB=0.4 GeV$^2$\\
			\hline\hline
			$\jpsi$&194 & 126 & 97 \\
			\hline
			$\Upsilon$&297 & 222 & 213 \\
			\hline
			\hline
			$\Upsilon^{'}$&186 & 113 & 89 \\
			\hline
		\end{tabular}
		\caption{The dissociation temperature($T_D$) for the
			quarkonia states (in units of MeV), when $\Gamma$= 2BE at $\xi$ =0.3} 
		\label{tab:t2}
	\end{center}
\end{table} 
We can observe a slight increase in dissociation temperatures when there is no magnetic field, compared to situations where a magnetic field is present for both $\xi =0.1$ and $0.3$. As previously discussed, the table highlights that $\Upsilon$ dissociates at a higher temperature than $J/\psi$ and $\Upsilon^{'}$. It is notable that the dissociation temperature of quarkonia states diminishes with the increasing magnetic field. This trend closely mirrors the reduction of the transition temperature due to the magnetic field, linked to the IMC effect. In essence, the consistent pattern of decreasing dissociation temperature with an increasing magnetic field persists in this scenario. Similar behavior was observed with slightly modified magnitude in our previous work ~\cite{Nilima:2022tmz} where we had considered the isotropic medium.

\section{Conclusions}
\label{sec4}

In summary, our investigation focused on the dissociation temperatures of charmonium and bottomonium states, considering momentum anisotropy in the medium. We revisited the medium-modified heavy quark-antiquark potential in the presence of a finite magnetic field, incorporating the temperature and magnetic field-dependent Debye mass within the gluon propagator. This inclusion allowed us to capture the essence of IMC through quark condensates inspired by LQCD. Examining the effects of different magnetic field strengths ($eB= 0, 0.2$ GeV$^2$, and $0.4$ GeV$^2$), we explored the real and imaginary parts of the potential at two temperatures, $T= 100 $ MeV and $T= 200$ MeV for $\xi=0.1$ and $0.3$. Our findings revealed an increasing screening mass with higher magnetic field strength, indicating enhanced screening at elevated temperatures, $T = 200$ MeV compared to lower temperatures, $T = 100$ MeV, both for parallel ($\theta_r =0$) and perpendicular ($\theta_r = \pi/2$) orientations.
Additionally, we observed an increment in the magnitude of the imaginary part of the potential with separation distance ($r$) for higher magnetic field values. Furthermore, for a given separation distance, the imaginary part exhibited greater magnitude at higher temperatures, $T = 200$ MeV, compared to lower temperatures, $T = 100$ MeV. Notably, the impact of anisotropy on the imaginary part was found to be less pronounced than on the real part. The observation is similar for both anisotropies ($\xi=0.1$ and $0.3$) considered.

Further insights were gained by studying the wavefunctions of $1s$ and $2s$ states for charmonium and bottomonium, revealing distinctions in peak height and width that contribute to our understanding of the binding nature of these states. Employing these wavefunctions, we obtained the binding energy and thermal dissociation width for various quarkonia states, and by employing a criterion where twice the binding energy equals the thermal dissociation width, we determined the dissociation temperatures for those states. Notably, we observed a decrease in the dissociation temperature of quarkonia states with an increase in magnetic field strength at fixed momentum anisotropy, $\xi=0.1$ and $\xi=0.3$. We also observed the slightly higher values of the dissociation temperature at $\xi=0.3$ for the quarkonium states at different magnetic field strengths as compared to $\xi=0.1$. Looking ahead, our future work aims to extend the present study by incorporating viscous corrections and examining the dissociation of heavy quarkonia in a hydrodynamically expanding viscous QGP.

\section*{Acknowledgement}
We would like to thank Aritra Bandyopadhyay for helpful discussions. IN acknowledges the Women Scientist Scheme A (WoS A) of the Department of Science and Technology (DST) for funding with grant no. DST/WoS-A/PM-79/2021. M.Y.J. would like to acknowledge the SERB-NPDF (National postdoctoral fellow) with File No. PDF/2022/001551.

\end{document}